\documentclass[12pt]{article}
\usepackage{amsmath,
graphicx, enumerate}
\usepackage[letterpaper,top=0.5in, bottom=1in, left=0.8in, right=0.8in, headsep=0.in, centering]{geometry}
\usepackage{amssymb}
\usepackage{epsfig}
\usepackage[normalem]{ulem}

\usepackage{soul}

\title{Wave-Particle Duality and the Hamilton-Jacobi Equation
}

\author{Gregory I. Sivashinsky\\  
Sackler Faculty of Exact Sciences,
School of Mathematical Sciences, \\
Tel Aviv University, Tel Aviv 69978, Israel}
\date{}
\begin{document}
\maketitle
\begin{center}
{\bf Abstract} 
\end{center}
\noindent The Hamilton-Jacobi equation of relativistic quantum mechanics is revisited.  The equation 
is shown to permit
 solutions in the form of breathers (oscillating/spinning solitons), displaying simultaneous particle-like and wave-like behavior.  The de Broglie wave thus acquires a clear {\it deterministic} meaning of a wave-like excitation of the classical action function.\\
\noindent The  problem of quantization in terms of the breathing action function and the double-slit experiment are discussed. 

\bigskip

\noindent {\bf PACS}: 03.65-w, ~03.65Pm, ~ 03.65.-b

\medskip

\noindent {\bf  Key words}:   de Broglie waves; wave-particle duality; relativistic wave equation

\setcounter{equation}{0}

\bigskip

\bigskip

\bigskip
\noindent{\bf 1. Introduction}\\
\noindent  A mathematical representation of the dual wave particle nature of matter remains one of the major challenges of quantum theory [1-7].  The present  study is an attempt to resolve this issue through an appropriately revised Hamilton-Jacobi formalism.\\ 
\noindent Consider the relativistic Hamilton-Jacobi (HJ) equation for a particle in an electromagnetic field,

\begin{eqnarray}\label{1}
\left(1/c^2\right)\left(\partial S/\partial t+e U\right)^2-\left(\nabla S-e {\mathbf A}/c\right)^2=m^2c^2
\end{eqnarray}

Here $U , {\mathbf A}$ are scalar and vector potentials of the field obeying the Lorentz calibration condition,
\begin{eqnarray}\label{2}
\left(1/c\right) \partial U/ \partial t  +  {\rm div} {\mathbf A} = 0
\end{eqnarray}

\noindent  The trajectory ${\mathbf x}(t)$ of the particle is governed by the equation [1],
\begin{eqnarray}\label{3}
\frac{d {\mathbf x}}{dt} = - \frac{c^2\left(\nabla S^{(0)}-e {\mathbf A}/c\right)}{\partial S^{(0)}/\partial t+e U} \quad ,
\end{eqnarray}

\noindent where $S^{(0)}$ is an appropriate action function of the system.
\noindent  As  is well known, the trajectories ${\mathbf x}(t)$ are {\it characteristics} of Eq.\:(1) [8], which are, in turn, the {\it traces} of small perturbations.  Indeed, let's represent the action function as
\begin{eqnarray}\label{4}
S=S^{(0)}+s
\end{eqnarray}

\noindent where $s$ is a perturbation.  Assuming $s$ to be small, Eq.\:(1) yields the linear equation,
\begin{eqnarray}\label{5}
\left(1/c^{2}\right) \left(\partial S^{(0)}/ \partial t + e U \right) \partial s / \partial t - \left(\nabla S^{(0)}-e {\mathbf A} /c \right) \nabla s = 0 \quad ,
\end{eqnarray}

\noindent whose characteristics are  identical to those of Eq.\:(1).\\
\noindent Let
\begin{eqnarray}\label{6}
{\mathbf f} \left({\mathbf x}, t \right) = {\mathbf c} 
\end{eqnarray}

\noindent be a set of independent integrals of Eq.\:(3), $\mathbf c$ being a constant vector.  Then the general solution of Eq.\:(5) may be written as,
\begin{eqnarray}\label{7}
s=s\left[{\mathbf f} \left({\mathbf x}, t \right) \right]
\end{eqnarray}

\noindent  The perturbation $s$ is, therefore, advected along the trajectory ${\mathbf x} (t)$ with the velocity ${\mathbf v}=d{\mathbf x}/d t$.
\noindent  If the perturbation is localized enough it will mimic the motion of the particle.  
\noindent As an example, consider the case of a free particle $(U=0, {\mathbf A}=0)$, where
\begin{eqnarray}\label{8}
S^{(0)}=-Et+ {\mathbf p} {\mathbf \cdot}{\mathbf x} \qquad \left(E^{2}/c^{2}=p^{2}+m^{2}c^{2}\right)
\end{eqnarray}

\noindent  Eq.\:(5) then yields,
\begin{eqnarray}\label{9}
s=s\left({\mathbf x} - {\mathbf v} t \right),
\end{eqnarray}

\noindent where
\begin{eqnarray}\label{10}
{\mathbf v}=c^2 {\mathbf p}/E
\end{eqnarray}

\noindent  In the linear approximation the perturbation is advected without changing its shape.  However, in a nonlinear description, due to the Huygens principle, the perturbation will gradually decay thereby implying stability (albeit nonlinear) of the regular solution $S^{(0)}$.\\
\noindent  The question is whether it is possible to modify the HJ equation (1) so that the new equation would allow for localized, nonspreading and nondecaying perturbations (excitations) of the regular action function.  Moreover, if the localized excitation breathes (oscillates/spins), one would end up with a {\it deterministic} model for a particle with quantum-like features.  As we intend to show, this kind of behavior can be successfully modeled by the conventional quantum Hamilton-Jacobi (QHJ) equation,
\begin{eqnarray}\label{11}
\left(1/c^2\right)\left(\partial S/\partial t+e U\right)^2 - \left(\nabla S- e {\mathbf A} /c\right)^2 = m^2c^2+i \hbar \Box S \quad ,
\end{eqnarray}

\noindent whose capacity, it transpires, has simply not been fully explored.
\newpage

\noindent {\bf 2. A free particle} \\
\noindent As  is well known, the QHJ equation (11) is a transformed version of the linear Klein-Gordon (KG) equation
\begin{eqnarray}\label{12}
\left(1/c^2\right) \left(\partial /\partial t + i e U /\hbar \right)^2 \Psi - \left(\nabla - i e {\mathbf A}/c \hbar \right) ^2 \Psi + \left(mc/\hbar \right)^2 \Psi=0,
\end{eqnarray}

\noindent obtained through the substitution,
\begin{eqnarray}\label{13}
\Psi= \exp \left(i S / \hbar \right)
\end{eqnarray}

\noindent  Consider first the case of a free particle  $(U=0,{\mathbf A}= 0)$ where Eq.\:(12) becomes
\begin{eqnarray}\label{14}
\Box \Psi+(mc/\hbar)^2 \Psi=0 \qquad (\Box=(1/c^2) \partial^2/\partial t^2 - \nabla^2)
\end{eqnarray}

\noindent  The KG equation (14) allows for a two-term spherically symmetric solution
\begin{eqnarray}\label{15}
\Psi= \exp [-i(mc^2/\hbar)t]+ \alpha \exp (-i \omega t)j_0(kr)
\end{eqnarray}

\noindent where
\begin{eqnarray}\label{16}
\omega = c \sqrt{k^2+(mc/\hbar)^2},
\end{eqnarray}

\noindent $r=\sqrt{x^2+y^2+z^2}$, $\alpha $ is a free parameter, and
\begin{eqnarray}\label{17}
j_0(kr)= \sin (kr)/kr
\end{eqnarray}

\noindent is the zeroth-order spherical Bessel function.

\noindent  The second term in Eq.\:(15) is a standing spherically symmetric breather, $\left|\alpha \right|$ being its intensity.
\noindent  In terms of the action function $S$, by virtue of (13), Eq.\:(15) readily yields
\begin{eqnarray}\label{18}
S=mc^2t-i \hbar \ln \left\{1+ \alpha \exp [-i( \omega - mc^2/\hbar)t]j_0(kr)\right\}
\end{eqnarray}

\noindent  Here the first term corresponds to the classical action function,  $S^{(0)}=-mc^2t$, for a free particle in the rest system while the second term represents its localized excitation, oscillating and {\it nonspreading}.  
 
 \noindent  Let's set the frequency of oscillations in Eq.\:(18) in accordance with the de Broglie postulate that each particle at rest can be linked to an internal `clock' of frequency $mc^2/\hbar$.
 \noindent  The frequency $\omega$ in Eq.\:(15) should therefore be specified as
 \begin{eqnarray}\label{19}
 \omega=2(mc^2/\hbar)
 \end{eqnarray}
 
 \noindent  Hence, by virtue of Eq.\:(16),
 \begin{eqnarray}\label{20}
 k=\sqrt{3}(mc/\hbar)
 \end{eqnarray}
 
 \noindent Eq.\:(18) thus becomes,
 \begin{eqnarray}\label{21}
 S= - mc^2t-i\hbar \ln \left\{1+ \alpha \exp \left[-i \left(\frac{mc^2}{\hbar}\right) t \right] j_0 \left[\sqrt{3}\left(\frac{mc}{\hbar}\right)r \right]\right\}
 \end{eqnarray}

\bigskip 
\noindent Note that in Eq.\:(21) the frequency is not affected by the nonlinearity of the system, preserving its value irrespective of the breather intensity.

\noindent Away from the breather's core ($r \gg\hbar/mc$ ),
\begin{eqnarray}\label{22}
S=-mc^2t-i\alpha\hbar\exp\left[-i\left(\frac{mc^2}{\hbar}\right)t\right] j_0 \left[\sqrt{3}\left(\frac{mc}{\hbar} \right) r \right]
\end{eqnarray}

\noindent The oscillations are therefore asymptotically {\it monochromatic}, again in accord with the de Broglie picture [1].

\noindent Similar to oscillations of an ideal pendulum,  the breather (21) is stable to small perturbations. The stability follows from the linearity of the KG equation.  Due to the linearity there is no coupling between the basic solution (15) and its perturbation, which also obeys the KG equation.  Therefore, if the initial perturbation is small it will remain so indefinitely.  The stability here is understood in a weak (non-asymptotic) sense. \\
\noindent  Physically relevant action functions are quite specific {\it global} solutions defined over the entire time axis  $-\infty<t<\infty$.  Moreover, they may be multiple valued and bound in space.  Such solutions cannot be obtained through a conventional initial-value problem unless suitable initial conditions are known in advance.

\noindent Until now we have dealt with a particle at rest.  For a particle moving at a constant velocity         $v$     along, say,        $x$  - axis, the corresponding expression for the action function is readily obtained from Eq.\:(21) through the Lorentz transformation,
\begin{eqnarray} \label{23}
t\to \frac{t-xv/c^2}{\sqrt{1-(v/c)^2}}\;, \qquad x\to\frac{ x-vt}{\sqrt{1-(v/c)^2}}
\end{eqnarray}

\noindent The transformed Bessel function      $j_0$          will then mimic the motion of the classical particle while the transformed temporal factor $\exp\left[-i (mc^2/\hbar)t\right]$ will turn into the associated 
de Broglie wave, thereby demonstrating {\it simultaneous} particle-like and wave-like behavior.   Moreover, unlike conventional quantum mechanics, here the modulated de Broglie wave acquires the clear {\it deterministic} meaning of a wave-like excitation of the action function, a complex-valued potential in configuration space.


\noindent If, as is conventional, we associate the gradients    $-\partial S/\partial t, \nabla S$                            
with the particle energy         $E$        and momentum      ${\bf p}$, then the Einstein relation
$(1/c^2)E^2=p^2+m^2c^2$
appears to hold only   far from the     $\hbar/mc$ - wide breather's core, or {\it on average} over the entire breather.  
The correspondence with classical relativistic mechanics  is therefore complied with. 




\noindent In addition to spherically symmetric breathers, Eq.\:(14) also permits  asymmetric breathers, spinning around  some axis.  In the latter case the second term of Eq.(15) should be replaced by
\begin{eqnarray}\label{24}
\alpha\exp\left[-2i\left(\frac{mc^2}{\hbar}\right)t+in\phi\right]j_l\left[\sqrt{3}\left(\frac{mc}{\hbar}\right)r\right]P_l^n\left(\cos\theta\right),
\end{eqnarray}							
where       $j_l$, $P_l^n$    are high-order spherical Bessel functions and associated Legendre functions.  It would be interesting to ascertain in what way (if any) the double-valued spin \!-$\tfrac{1}{2}$ breather 
may be linked to the Dirac wave function.\\  

\noindent The next question is how to reproduce quantization directly in terms of the breathing  action function.  The geometrically simplest situation, where such an effect manifests itself, is the periodic motion of an otherwise free particle over a closed interval   $0<x<d$.  In this case the field-free version of Eq.(11) must be considered jointly with two boundary conditions,
\begin{eqnarray}\label{25}
\partial S(0,y,z,t)/\partial t=\partial S(d,y,z,t)/\partial t \:, \nonumber \\
\partial S(0,y,z,t)/\partial x=\partial S(d,y,z,t)/\partial x
\end{eqnarray}

\noindent Any classical action function for a free particle,
\begin{eqnarray}\label{26}
 S=-Et+px  
\end{eqnarray}

\noindent is clearly a solution of this problem.  However, in the case of a breathing 
action function the situation proves to be different.  Thanks to the boundary conditions (25), the moving breather interacts with itself, and this may well lead to its self-destruction unless some particular conditions are met.                                         

\noindent Consider first the simplest case of a particle at rest $(v=0)$.  The pertinent solution is readily obtained by converting the problem for a finite interval into a problem for an infinite interval ($-\infty<x<\infty$) filled with a      $d$-periodic train of standing breathers, assumed to be spherical for simplicity.  The resulting action function then reads,
\begin{eqnarray}\label{27}
S=-mc^2t-i\hbar\ln\left\{ 1+\alpha\exp\left[-i\left(\frac{mc^2}{\hbar}\right)t\right]\sum_{k}\left( j_0\right)_{d,0}^{(k)}\right\},
\end{eqnarray}

\noindent where 										
\begin{eqnarray}\label{28}
\left(j_0\right)_{d,0}^{(k)}=\frac{\sin \left[\sqrt{3}\left(mc/\hbar\right)r_{d,0}^{(k)}\right]}{\sqrt{3}\left(mc/\hbar\right)r_{d,0}^{(k)}}\;,
\end{eqnarray}
\begin{eqnarray}\label{29}
r_{d,0}^{(k)}=\sqrt{(x-kd)^2+y^2+z^2} \qquad (k=0,\pm 1,\pm 2,...)
\end{eqnarray}
Here the second subscript stands for    $v=0$.     

\noindent The action function for a moving particle $(v\ne0)$  is obtained from (27) (28) (29) through the Lorentz transformation (23), provided  $d$  is replaced by   $d/\sqrt{1-(v/c)^2}$. The latter step is needed to balance the relativistic contraction, and thereby to preserve the spatial period ($d$)  of the system.  The resulting action-function thus becomes, 
\begin{eqnarray}\label{30}
  S=-Et+px-i\hbar\ln \left\{1+\alpha\exp\left[i\left(\frac{-Et+px}{\hbar}\right)\right]\sum_{k}\left(j_0\right)_{d,v}^{(k)}\right\},         
\end{eqnarray}
where
\begin{eqnarray}\label{31}
\left(j_0\right)_{d,v}^{(k)}=\frac{\sin\left[\sqrt{3}\left(mc/\hbar\right)r_{d,v}^{(k)}\right]}{\sqrt{3}\left(mc/\hbar\right)r_{d,v}^{(k)}}\;,
\end{eqnarray}
\begin{eqnarray}\label{32}
r_{d,v}^{(k)}=\sqrt{\left(\frac{x-vt-kd}{\sqrt{1-(v/c)^2}}\right)^2+y^2+z^2}
\end{eqnarray}
The spatial 	$2\pi\hbar/p$ - periodicity of  $\exp\left[i\left(-Et+px\right)/\hbar\right]$								
is compatible with the spatial     $d$   - periodicity of      
$\sum_{k}\left(j_0\right)_{d,v}^{(k)}$
 only if
\begin{eqnarray}\label{33}
dp=2\pi n\hbar \qquad (n=0,1,2,3,...)\;,
\end{eqnarray}
which recovers the familiar Bohr-Sommerfeld quantum  condition.  While the particle velocity is clearly subluminal its communication with boundary conditions is {\it superluminal}, which does not violate the Lorenz-invariance of the system.\\
\noindent The above solution (30)-(33) may be easily adapted for the problem of a particle shuttling between two perfectly reflecting walls, $x=0$ and $x=d/2$.  To handle the double-valuedness of the pertinent action function the trajectory of the particle, following 
the Einstein-Keller topological approach [9,10], should be placed on the double-sheeted strip, $0<x<d/2$, $-\infty<y<\infty$, $z=\pm0$.  Thereupon the problem reduces to the previous one.

\noindent{\bf 3. A particle in a slowly varying field}\\
\noindent  The de Broglie postulate holds at least for breathers exposed to slowly varying potentials, characterized by spatio-temporal scales much larger than $\hbar/mc, \hbar/mc^2$.
\noindent  Indeed, as may be readily shown, for slowly varying potentials, Eq.\:(21) becomes

\begin{eqnarray}\label{34}
S=-\left(mc^2+eU\right)t+ \frac{e}{c} \mathbf A \mathbf \cdot \mathbf x - i \hbar \ln \left\{1+ \alpha \exp \left[-i \left( \frac{mc^2}{\hbar} \right) t \right] j_0 \left[\sqrt{3}\left( \frac{mc}{\hbar} \right) r \right]\right\}
\end{eqnarray}

\noindent  So, unlike the action function as a whole, the frequency of its oscillations in the rest system is not affected by the field.  This invariance is untenable for the wave function $\Psi$ (13), which therefore cannot serve as a physically objective representation of the de Broglie clock.\\
\noindent For a particle {\it moving} in a slowly varying field the `fast' spatio-temporal coordinates $\mathbf x,t$ in Eq.\:(34) should be subjected to the Lorentz transformation, with the velocity $\mathbf v$ regarded as a slowly varying vector.

\noindent  The action function (34) and its Lorentz transformed version pertain to the interior of the breather (inner solution).  Away from the breather's core the action function is described by the regular (breather-free) solution of the QJH equation $S^{(0)}$, involving only large spatio-temporal scales (outer solution).  The inner solution is clearly affected by the outer solution through the velocity field $\mathbf v$, while the reverse influence does not take place, at least not for the leading order asymptotics.  The simplest picture emerges in the nonrelativistic {\it semiclassical} limit [11] where the uniformly valid asymptotic solution may be represented as,

\begin{eqnarray}\label{35}
S=-mc^2t+S_{sc}-i\hbar \ln \left\{1+ \alpha \exp \left[-i \left(\frac{mc^2+mv^2/2}{\hbar}\right)t\right]\exp \left[i \left(\frac{\mathbf p \mathbf \cdot\mathbf x}{\hbar} \right)\right]j_0 \left[\sqrt{3}\left(\frac{mc}{\hbar}\right)r \right]\right\}
\end{eqnarray}

\noindent  Here $S_{sc}$ is the semiclassical action function governed by the equation
\begin{eqnarray}\label{36}
\frac{\partial S_{sc}}{\partial t} + \frac{1}{2m} \left(\nabla S_{sc}-\frac{e}{c} \mathbf A\right)^{2}+eU=- \frac{i \hbar} {m} \nabla^2 S_c \quad ,
\end{eqnarray}

\noindent where $S_c$ is the classical action function obeying the equation
\begin{eqnarray}\label{37}
\frac{\partial S_c}{\partial t} + \frac{1}{2m} \left(\nabla S_c - \frac{e}{c} \mathbf A \right)^{2}+eU=0
\end{eqnarray}
The argument $r$ in $j_0\left[\sqrt{3}\left(mc/\hbar \right) r \right]$ is defined as
\begin{eqnarray}\label{38}
r= \left| \mathbf x - \mathbf x_p \left(t \right)\right|\quad ,
\end{eqnarray}
where  $\mathbf x_p\left(t\right)$ is the classical trajectory of the particle described by the equation
\begin{eqnarray}\label{39}
\frac{d \mathbf x _{p}}{dt} = \frac{1}{m} \mathbf p - \frac{e}{c} \mathbf A \qquad \left(\: \mathbf p = \nabla S_{c}\: \right )
\end{eqnarray}
\noindent One therefore may readily see the connection between the new formalism and those of Schr$\ddot{\mathrm{o}}$dinger, Bohr and Sommerfeld.  The Schr$\ddot{\mathrm{o}}$dinger formalism pertains to the outer solution which, under appropriate conditions, provides the data on the particle's range of energies, but says nothing about its trajectory.  The information about the particle's trajectory comes from the inner solution, connecting de Broglie waves with the Bohr-Sommerfeld theory.  Recall that for bound systems in the semiclassical limit (high quantum numbers), the Schr$\ddot{\mathrm{o}}$dinger and Bohr-Sommerfeld formalisms lead to identical energy spectra.\\

\bigskip
\noindent {\bf 4. The effect of boundaries}\\
The outlined formalism seems fully compatible with the double-slit experiment.  Diffraction pictures obtained from electron beams of very low intensity [1,4] provide convincing evidence that the double-slit experiment is actually a one-particle effect where the particle communicates with distant boundaries that affect its trajectory.  The breather passing through a slit `feels' whether the other slit is open or closed, and changes its trajectory accordingly.  A geometrically simpler system, where one may readily observe the impact of boundaries, is a breather moving in a uniform {\it circular} motion through a thin toroidal duct, e.g. doughnut shaped carbon nanotube.  In classical mechanics such a motion would be impossible without external forcing.  In quantum mechanics however, the bending of the trajectory is a manifestation of the rotating nature of the pertinent KG solution.  The problem may be easily tackled analytically in the intermediate limit when the width of the duct $d$ is comparable to the de Broglie wavelength, small compared to the torus centerline radius $R$, and large compared to the breather's width $\hbar / mc$.


\noindent For the leading order asymptotics the resulting action function (inner solution), written in cylindrical co-ordinates $\rho,\phi,z$, reads,
\begin{eqnarray} \label{40}
 S=-Et+p_{\phi} R\phi-i\hbar\ln \left\{1+\alpha\exp \left[ i\left(\frac{-Et+p_{\phi}R\phi}{\hbar}\right)\right]j_0 \left[\sqrt{3}\left( \frac {mc}{\hbar} \right)r \right]\right\},
 \end{eqnarray}
 
 \noindent  where 
 \begin{eqnarray}\label{41}
 r=\sqrt{\left(R\phi-v_\phi t \right)^2 + \left(\rho-R\right)^2 + z^2} \quad ,\quad E=mc^2+mv^2_{\phi}/2 \quad,\quad p_\phi=mv_\phi=n \hbar/R
 \end{eqnarray}
\noindent Here $v_\phi, p_\phi$ are the azimuthal velocity and momentum, $v_\phi \ll c \:, R \phi \sim \left| \rho -R\right|\sim \hbar/p_\phi, n\phi \sim 1,$ and $n$ is a large integer.\\

\bigskip
\noindent {\bf 5. Concluding remarks}\\
\noindent  The proposed formalism is certainly related to de Broglie's double solution program [1].  Yet, unlike the latter, in the current model the breather is guided by a regular, generally nonwaving, action function $S^{(0)}$ rather than by a guiding wave solution (Sec 2).  The guiding action function and its localized excitation (breather) are coupled through the {\it nonlinear} QHJ equation (11).  
  

\noindent The de Broglie's double solution program is adjacent to the de Broglie-Bohm pilot-wave theory [1,4,7], which should be seen as a degenerate double solution theory, where the breather has been replaced by the particle position governed by an appropriate guidance equation.  For unbound systems the pilot-wave theory and the current formulation are likely to correspond. However, for bound systems the de Broglie-Bohm formalism, based on standing KG-Schr$\ddot{\mathrm{o}}$dinger waves, produce immobile particles [1,4], which contradicts the classical limit.  In the present formulation, dealing with drifting nonspreading breathers and multiple valued action functions, this difficulty does not occur (Sec. 2).\\
\noindent At this stage it is difficult to see whether the amended QHJ formalism is indeed adequate enough to reproduce all the basic features of quantum-mechanical phenomenology.  In any case, a few preliminary observations already show that a mathematical representation of unified wave-particle behavior is quite feasible, even within the framework of the conventional QHJ equation.
\bigskip
\bigskip

\noindent {\bf Acknowledgments}


\noindent The author wishes to thank Irina Brailovsky, Mark Azbel and Eugene Levich for interesting discussions, and Victor P. Maslov and Steven Weinberg for stimulating correspondence. 

\noindent These studies were supported in part by the Bauer-Neumann Chair in Applied Mathematics and Theoretical Mechanics, the US-Israel Binational Science Foundation (Grant 2006-151), and the Israel Science Foundation (Grant 32/09).

\end{document}